\documentclass{elsart}

 \usepackage{graphicx}

\usepackage{amssymb,amsmath}
\usepackage{bm}
\usepackage{dcolumn}

\begin{document}

\begin{frontmatter}



\title{Monte Carlo studies of triangulated spherical surfaces in the two-dimensional space}


\author{Hiroshi Koibuchi}
\ead{koibuchi@mech.ibaraki-ct.ac.jp}

\address{Department of Mechanical and Systems Engineering, Ibaraki National College of Technology, 
Nakane 866, Hitachinaka,  Ibaraki 312-8508, Japan}

\begin{abstract}
We numerically study a triangulated surface model in ${\bf R}^2$ by taking into account a viewpoint of string model. The models are defined by a mapping $X$ from a two-dimensional surface $M$ to ${\bf R}^2$, where the mapping $X$ and the metric $g$ of $M$ are the dynamical variables. The sum over $g$ in the partition function is simulated by the sum over bond lengths and deficit angles by using the Regge calculus technique, and the sum over $g$ is defined to be performed independently of the sum over $X$. We find that the model undergoes a first-order transition of surface fluctuations, which accompanies a collapsing transition, and that the transitions are reflected in the internal geometry of surface. Fluid surface models are also studied on dynamically triangulated surfaces, and the transitions are found to be of second order. The order of the transition remains unchanged from that of the conventional model defined only by the variable $X$ both in the fixed-connectivity and the fluid models.
\end{abstract}

\end{frontmatter}

\section{Introduction}\label{intro}
Two-dimensional surfaces are interesting objects in the sense that the length scales are ranging from microscopic scales to macroscopic ones. The microscopic string \cite{POLYAKOV-PLB1981} and the macroscopic biological membranes \cite{NELSON-SMMS2004,Gompper-Schick-PTC-1994,Bowick-PREP2001,WIESE-PTCP19-2000} can be described and hence unified by a surface model of Helfrich and Polyakov \cite{HELFRICH-1973,POLYAKOV-NPB1986}.

The models are defined by a curvature Hamiltonian, which is given by an extrinsic curvature energy \cite{HELFRICH-1973,POLYAKOV-NPB1986,KLEINERT-PLB1986}. The surface shape or its motion is considered to be governed by the curvature Hamiltonian. One interesting phenomenon is a phase transition of shape transformation, which separates a flat phase at large bending rigidity from a crumpled phase at small bending rigidity \cite{P-L-1985PRL,David-1986EPL,DavidGuitter-1988EPL,BKS-2000PLA,BK-2001EPL,Kownacki-Mouhanna-2009PRE}. Numerical studies have been devoted to understand the phase transitions \cite{KANTOR-NELSON-PRA1987,CATTERALL-NPBSUP1991,AMBJORN-NPB1993}. While recent simulations show that the model undergoes a first-order transition \cite{KD-PRE2002,NISHIYAMA-PRE-2004,KOIB-PRE-2004,KOIB-PRE-2005,KOIB-NPB-2006}, a continuous transition is also predicted by theoretical studies based on the renormalization group technique \cite{DavidGuitter-1988EPL,Kownacki-Mouhanna-2009PRE}. Thus, the transition still remains to be studied. 

In the string model context, a two-dimensional surface $M$ and the image $X(M)$ of the mapping $X$ from $M$ to ${\bf R}^d$ are independently treated. The length scale of $M$ is not always identified to the one in $X(M)$. This is because the mapping $X$ and the metric $g$ of $M$ are considered as the two different dynamical variables. To the contrary, in the case of numerical studies for real membranes, the two-dimensional surface $M$ is always fixed and hence is not taken into account to define the model. This corresponds to the case where the Euclidean flat metric $g_{ab}\!=\!\delta_{ab}$ is assumed in $M$. In this case, the length scale in $M$ is fixed; the bond length of the triangulated $M$ is fixed to some constant. In the case of the induced metric $g_{ab}\!=\!\partial_a X^\mu \partial_b X^\mu$, $M$ can be identified with $X(M)$. In this case, the length scale of $M$ is identified to that of the external space ${\bf R}^3$, which is also an Euclidean flat space. The surface position $X(M)$ is the only dynamical variable in both cases. Thus, little is known about the dependence of the phase structure on the variable $g$ in the triangulated surface models. Therefore, it is still interesting to study numerically the surface model described by both of the variables $X$ and $g$ and to see whether the phase structure of the model depends on $g$ or not. 

There have been a lot of studies on models defined by a mapping from a $D$-dimensional surface $M$ to the $d$-dimensional space ${\bf R}^d$, including $d\!=\!0$ the matrix model \cite{FGZ-PR-1995}, from the viewpoint of string model \cite{GinspargMoore-TASI1992}, where the mapping $X$, including $M$, is considered to be dynamically generated. However, the phase structure of surface models has not been so extensively studied at least numerically by assuming the metric as a dynamical variable in the low-dimensional cases $d\!=\!2,3$. 

In this paper, we numerically study a surface model defined by a mapping $X$ from $M$ to ${\bf R}^2$, where the simulation is computationally less time-consuming than ${\bf R}^3$. The variable $g$ is summed over by using the Regge calculus technique \cite{REGGE-NC1961,HAMBER-LH1986,FDAVID-LH1992} in the partition function. $M$ is topologically fixed as a sphere, and as a consequence the sum over topology is not included in the sum over $g$. 

This paper is organized as follows: in Section \ref{model} the model is defined on both fixed-connectivity surfaces and dynamically triangulated surfaces. In Section \ref{continuous-model} we describe how to obtain the model from the continuous Hamiltonian for strings. A discrete metric tensor is introduced to obtain the discrete model. In Section \ref{MC_technique} detailed information of Monte Carlo (MC) technique is given. The numerical results including those of the conventional models are presented in Section \ref{Results}, and we summarize the results in the final section \ref{Conclusion}. 

\section{Model}\label{model}
\begin{figure}[h!]
\centering
\includegraphics[width=10.5cm]{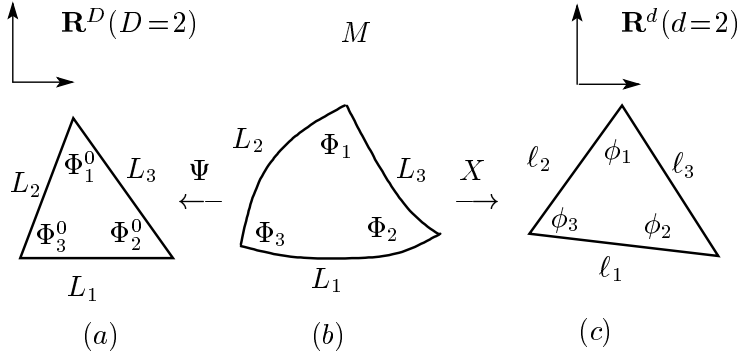}  
\caption{(a) The tangential plane $\Psi({\it \Delta})$ of (b) a triangle ${\it \Delta}$ in $M$, and (c) the image $X({\it \Delta})$ in ${\bf R}^d\left(d\!=\!2\right)$, where $\Psi$ is a coordinate mapping from ${\it \Delta}$ to ${\bf R}^D(D\!=\!2)$, and  $X$ is a mapping from $M$ to ${\bf R}^d\left(d\!=\!2\right)$. The triangle ${\it \Delta}$ in $M$ is almost flat but slightly curved, while $\Psi({\it \Delta})$ and $X({\it \Delta})$ are assumed to be linear.} 
\label{fig-1}
\end{figure}
Let $M$ be a two-dimensional spherical surface, which is not always included in the three-dimensional Euclidean space ${\bf R}^3$, and $M$ is assumed to be triangulated with smooth triangles. The reason why $M$ is triangulated is because the numerical studies including the one in this paper are always performed on triangulated surfaces. The triangulation is characterized by the three numbers $N$, $N_B(=\!3N\!-\!6)$, and $N_T(=\!2N\!-\!4)$, which are the total number of vertices, the total number of bonds, and the total number of triangles, respectively.  A linear triangle in ${\bf R}^D(D\!=\!2)$ corresponds to a triangle in $M$ by a coordinate mapping $\Psi$; the triangle edges are considered as local coordinate axes on $M$. $M$ is also assumed to be embedded in the external space ${\bf R}^d\left(d\!=\!2\right)$ as $X(M)$ by a mapping $X$. Fig.\ref{fig-1} shows the linear triangle $\Psi({\it \Delta})$ in ${\bf R}^D(D\!=\!2)$, the triangle ${\it \Delta}$ in $M$, and the image $X({\it \Delta})$, which is also a linear triangle in ${\bf R}^d\left(d\!=\!2\right)$. If $M$ is included in ${\bf R}^d(d\!=\!2)$, the triangle ${\it \Delta}$ in Fig.\ref{fig-1}(b) and the tangential triangle $\Psi({\it \Delta})$ in Fig. \ref{fig-1}(a) should be identified with the triangle $X({\it \Delta})$ in Fig. \ref{fig-1}(c). 

\begin{figure}[htb]
\centering
\includegraphics[width=10.5cm]{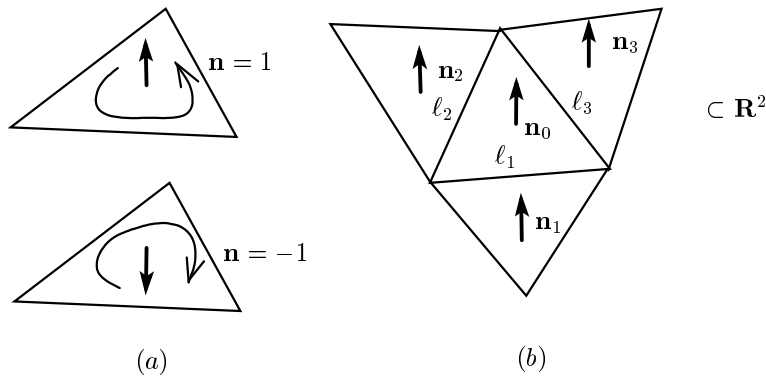}
\caption{(a) Two possible values of the unit normal vector ${\bf n}$ of the triangle $X({\it \Delta})$, and (b) the unit normal vectors ${\bf n}_i$ $(i\!=\!1,2,3)$ of the nearest neighbor triangles.} 
\label{fig-2}
\end{figure}

Our basic assumption is that $M$ is not always included in ${\bf R}^d(d\!=\!2)$. In order to define a length scale in $M$, we identify the edge length $L$ of ${\it \Delta}$ with that of $\Psi({\it \Delta})$. It is not unreasonable to identify $L$ of ${\it \Delta}$ with that of $\Psi({\it \Delta})$ because the length scale in $M$ can locally be fixed by a coordinate mapping $\Psi$; where the triangle surfaces are not always identified between ${\it \Delta}$ and $\Psi({\it \Delta})$. Thus, the triangle inequalities 
\begin{equation} 
\label{triangle-rel}
 L_i+L_j>L_k 
\end{equation} 
are satisfied on ${\it \Delta}$ in $M$. Only difference between ${\it \Delta}$ and $\Psi({\it \Delta})$ is in the internal angles and in the area $A_{\it \Delta}$ of ${\it \Delta}$, where $A_{\it \Delta}$ is defined by
\begin{equation} 
\label{area}
A_{\it \Delta}=\frac{1}{6}\left(L_1L_2|\sin\Phi_3|+L_2L_3|\sin\Phi_1|+L_3L_1|\sin\Phi_2|\right),
\end{equation} 
where $|\sin\Phi_i|$ is because $\Phi_i$ is not always constrained to be $0\!<\!\Phi_i\!<\!\pi$. While the sum of internal angles of the linear triangles is given by $\sum_{i=1}^3\Phi^0_i\!=\!\pi$ ($0\!<\!\Phi^0_i\!<\!\pi$), and $\sum_{i=1}^3\phi_i\!=\!\pi$ ($0\!<\!\phi_i\!<\!\pi$), the sum $\sum_{i=1}^3\Phi_i$ is not always identical to $\pi$ on ${\it \Delta}$. The condition $0\!<\!\Phi_i\!<\!\pi$ corresponds to the ordinary triangles such as the one shown in Fig. \ref{fig-2}(b), and moreover $\Phi_i\left(\notin [0,\pi]\right)$ also corresponds to those triangles since $|\sin\Phi_i|$ in Eq.(\ref{area}) is defined to represent the area of those triangles. 

The deficit angle ${\it \varphi}$ of ${\it \Delta}$ is defined by
\begin{equation} 
\label{deficita}
\varphi=\sum_{i=1}^3 \Phi_i\!-\!\pi.
\end{equation}

Therefore, the internal angle $\Phi_i$ of ${\it \Delta}$ can be expressed by
\begin{equation} 
\label{internal_angle}
 \Phi_i=\Phi^0_i\left(1+\frac {\varphi}{\pi}\right),\quad(i=1,2,3). 
\end{equation} 
We should note that $\Phi^0_i$ is given only by the edge length $L$ on $\Psi({\it \Delta})$. Since $L$ is identified to the one of ${\it \Delta}$, $\Phi_i$ is given by using only $L$ and $\varphi$. Thus, the sum over $L$ and $\varphi$ can simulate the sum over metric $g$ on $M$, where ${\it \varphi}$ and $L$ are assumed to be independent of each other on ${\it \Delta}$. This is a Regge calculus approach to the sum over metrics $g$ on $M$ \cite{FDAVID-LH1992}.

It is possible that  $\lim_{N\!\to\!\infty}(1/N_T)\sum_{i=1}^{N_T} \varphi_i\!=\!0$ is violated:
\begin{equation}
\label{sum_deficit}
\lim_{N\!\to\!\infty} \frac{1}{N_T}\sum_{i=1}^{N_T} \varphi_i\not= 0.
\end{equation}
The reason of this is as follows: $\sum_{i=1}^{N_T}\varphi_i$ is not always identical to the sum over deficit angles $\sum_{i=1}^N\delta_i$, where $\delta_i$ is the deficit angle defined by $\delta_i\!=\!2\pi\!-\!\sum_{j(i)}\Phi_{j(i)}$, where $\Phi_{j(i)}$ is an internal angle of the triangle $j(i)$ meeting at the vertex $i$. In fact, if $M$ is piece-wise linearly triangulated, we have $\sum_{i=1}^N\delta_i\!=\!2\pi\chi \!=\!4\pi$ while $\sum_{i=1}^{N_T}\varphi_i\!=\!0$, where $\chi$ is the Euler number. On the contrary, we assume that $M$ is smoothly triangulated, where "smoothly triangulated" means that every ${\it \Delta}$ is a smooth triangle and $\delta_i\!=\!0$ at every vertex $i$.  In this case we have $\sum_{i=1}^N\delta_i\!=\!0$ while $\sum_{i=1}^{N_T}\varphi_i\!=\!4\pi$, which is the prediction of the Gauss-Bonnet theorem $\int\sqrt{g} d^2x K\!=\!2\pi\chi \!=\!4\pi$ on smoothly triangulated $M$, where $K$ is the Gaussian curvature. Therefore $(1/N_T)\sum_{i=1}^{N_T}\varphi_i\!=\!0$ $(N\!\to\!\infty)$ is satisfied if $M$ is piece-wise linearly triangulated or smoothly triangulated. However, the above mentioned basic assumption for $M$ does not always imply that $M$ is either linearly triangulated or smoothly triangulated, because the constraints $\varphi_i\!=\!0$ and $\delta_i\!=\!0$ are not imposed on the triangulation. 

We comment on the range of $\varphi$. A constraint on $\varphi(\in {\bf R})$ is given by the integration measure described below in this section. As a consequence, $|\varphi|$ is limited to have a value in some finite range in ${\bf R}$. Thus, the internal angles $\Phi_j$ of a triangle $i$ are automatically determined by Eq. (\ref{internal_angle}) from a given $\varphi_i$. Therefore $\Phi$ is not always constrained to be $0\!<\!\Phi\!<\!\pi$ as mentioned above. Informations on the value of $\varphi$, the variance of $\sum_{i=1}^{N_T}\varphi_i$, and the range of $\varphi$ are given in the final part of Section \ref{Results}.

Although the coordinate axes are assumed to be varied independently of the mapping $X$, the dynamical triangulation technique is still interesting from the viewpoint of reparametrization invariance in the discrete model. The model on dynamically triangulated fluid surfaces is also studied in this paper. 

The discrete Hamiltonian $S$ of the model on the triangulated surface $M$ is defined such that 
\begin{eqnarray}
\label{Disc-Eneg-model-12} 
&&S\left(X,\{L,\varphi\}\right)=S_1+bS_2, \nonumber \\
&&S_1=\frac{ 1}{ 12}\sum_{\it \Delta} S_1\left({\it \Delta}\right)/A_{\it \Delta},\quad
S_2=\frac{ 1}{ 12}\sum_{\it \Delta} S_2\left({\it \Delta}\right)/A_{\it \Delta}, 
\end{eqnarray}
where $S_1$ is the Gaussian bond potential, and $S_2$ is the bending energy. $A_{\it \Delta}$ in $S_1$ and $S_2$ is the area of the triangle ${\it \Delta}$ in $M$ and is defined by Eq.(\ref{area}). $S_1\left({\it \Delta}\right)$ in $S_1$ is defined by 
\begin{eqnarray}
\label{Disc-Eneg-S-1} 
S_1\left({\it \Delta}\right)=&&\ell_1^2\left(L_2^2+L_3^2\right)-2\ell_1\ell_2\cos\phi_3\;L_1L_2\cos\Phi_3   \nonumber \\
&&\!\!\!\!+\ell_2^2\left(L_3^2+L_1^2\right)-2\ell_2\ell_3\cos\phi_1\;L_2L_3\cos\Phi_1   \nonumber \\
&&\!\!\!\!+\ell_3^2\left(L_1^2+L_2^2\right)-2\ell_3\ell_1\cos\phi_2\;L_3L_1\cos\Phi_2,  
\end{eqnarray} 
where the symbols $L_i$, $\Phi_i$ are the bond length and the internal angles of ${\it \Delta}$ in $M$, and $\ell_i$, $\phi_i$ are those of $X({\it \Delta})$ in ${\bf R}^d(d\!=\!2)$. The symbol $S_2\left({\it \Delta}\right) $ in $S_2$ is given by
\begin{eqnarray}
\label{Disc-Eneg-S-2} 
S_2\left({\it \Delta}\right)=&& 2L_1^2\left(1\!-\!{\bf n}_0\!\cdot\!{\bf n}_1\right)
-L_1L_2\cos\Phi_3\;\left({\bf n}_1\!-\!{\bf n}_0\right)\!\cdot\!\left({\bf n}_2\!-\!{\bf n}_0\right)  \nonumber \\
&&\!\!\!\!+2L_2^2\left(1\!-\!{\bf n}_0\!\cdot\!{\bf n}_2\right)
-L_2L_3\cos\Phi_1\;\left({\bf n}_2\!-\!{\bf n}_0\right)\!\cdot\!\left({\bf n}_3\!-\!{\bf n}_0\right)  \nonumber \\
&&\!\!\!\!+2L_3^2\left(1\!-\!{\bf n}_0\!\cdot\!{\bf n}_3\right)
-L_3L_1\cos\Phi_2\;\left({\bf n}_3\!-\!{\bf n}_0\right)\!\cdot\!\left({\bf n}_1\!-\!{\bf n}_0\right), 
\end{eqnarray} 
where ${\bf n}_0\left(\in {\bf Z}_2\!=\!\{-1,1\}\right)$ is a unit normal vector of the triangle $X({\it \Delta})$, and ${\bf n}_i$ is a unit normal vector of the nearest neighbor triangle $i$. The value of ${\bf n}$ is naturally defined by the orientation of the triangle. Figure \ref{fig-2}(a) shows two different values of ${\bf n}$, and Fig. \ref{fig-2}(b) shows ${\bf n}_0$ and ${\bf n}_i$ $(i\!=\!1,2,3)$ corresponding to those in $S_2\left({\it \Delta}\right)$ of Eq.(\ref{Disc-Eneg-S-2}). We should note that ${\bf n}\notin {\bf R}^3$, and hence the model is defined within ${\bf R}^d(d\!=\!2)$.

The partition function of the model on fixed connectivity surfaces, which is denoted by {\it model 1}, is defined by
\begin{equation} 
\label{Part-Func}
 Z_{\rm fix}(b) =  \int [d L]\int [d \varphi]\int^\prime \prod _{i=1}^{N} d X_i \exp\left[-S(X,\{L,\varphi\})\right], \quad({\rm model \; 1}),
\end{equation} 
where the symbols $\int [d L]$ and $\int [d \varphi]$ are given by
\begin{equation} 
\label{measure_1}
\int [d L]=\int\prod _{i=1}^{N_B} d L_i \exp\left(-\sum_{i=1}^{N_B} L_i^2 \right), 
\end{equation} 
and
\begin{equation} 
\label{measure_2}
\int [d \varphi]=\int\prod _{i=1}^{N_T} d \varphi_i \exp\left(-\sum_{i=1}^{N_T} | \varphi_i | \right).
\end{equation} 
The symbol $L_i$ in Eq. (\ref{measure_1}) denotes the bond length of ${\it \Delta}$, and $\varphi_i$ in Eq. (\ref{measure_2}) denotes the deficit angle. The factors $\exp\left(-\sum_{i} L_i^2 \right)$ and $\exp\left(-\sum_i | \varphi_i | \right)$ are necessary in order to make the integrations of the variables $L \left(\in\!{\bf R}_{> 0}\right)$ and $\varphi \left(\in\!{\bf R}\right)$ well-defined; the factor $\exp\left(-\sum_{\it \Delta} A_{\it \Delta} \right)$ can also be used in place of $\exp\left(-\sum_{i} L_i^2 \right)$ in Eq. (\ref{measure_1}). It is possible to introduce parameters $\lambda_L$, $\lambda_\varphi$ in order to control the fluctuations of $L$ and $\varphi$ such that $\exp\left(-\lambda_L\sum_{i=1}^{N_B} L_i^2 \right)$ and $\exp\left(-\lambda_\varphi\sum_{i=1}^{N_T} | \varphi_i | \right)$, however, we assume $\lambda_L\!=\!1$ and $\lambda_\varphi\!=\!1$ for simplicity. The prime in $\int^\prime \prod _{i=1}^{N} d X_i$ in Eq. (\ref{Part-Func}) denotes that the center of mass of surface $X(M)$ is fixed. We note that $M$ is not always globally flat (or $M\!\subseteq\! {\bf R}^2$) even when $\Phi\!=\!\Phi_0$ because the deficit angles at the vertices are not always zero, this is because the edge length $L$ is integrated independently of the other edges.

The partition function of the model on dynamically triangulated surfaces, which is denoted by {\it model 2}, can be written by including the sum over possible triangulations $\sum_{\rm T}$ such that 
\begin{equation} 
\label{Part-Func-dyna}
 Z_{\rm flu}(b) =  \sum_{\rm T}\int [d L]\int [d \varphi]\int^\prime \prod _{i=1}^{N} d X_i \exp\left[-S(X,\{L,\varphi\})\right], \quad({\rm model \; 2}).
\end{equation} 
 $\sum_{\rm T}$ is performed by the bond flip technique. The bond flips is simultaneously performed both in $M$ and in $X(M)$, because the mapping $X$ preserves the triangulations. Since the edges of triangles in $M$ are considered as coordinate axes, the flip of bonds can make a large difference on the configuration and influence the equilibrium property. Thus, the bond flips should be carefully performed in model 2. The detailed information on this point is given in Section \ref{MC_technique}.

\section{Continuous model} \label{continuous-model}
We comment on a correspondence between the discrete model and the continuous model. The continuous energy $S_1$ is the Polyakov action for strings in ${\bf R}^d\left(d\!=\!2\right)$, and it is given by
\begin{equation} 
\label{cont_S1}
S_1=\int \sqrt{g}d^2x g^{ab} \partial_a X^\mu \partial_b X^\mu,
\end{equation} 
where $g$ is the determinant of the metric tensor $g_{ab}$, $\left(a,b\!=\!1,2\right)$ of $M$, and $g^{ab}$ is the inverse of $g_{ab}$. The symbol $\mu$ of $X^\mu$ denotes that $X^\mu \in {\bf R}^d\left(d\!=\!2\right)$. In order to obtain an explicit expression of $g_{ab}$, we consider the edges $L_1$ and $L_2$ of ${\it \Delta}$ in $M$ as the axes of a local coordinate. Thus, we define the discrete metric $g_{ab}$ such that
\begin{equation} 
\label{induced_metric}
g_{ab}=\left(  
       \begin{array}{@{\,}ll}
        L_1^2 & \; L_1L_2\cos\Phi_3 \\
       L_1L_2\cos\Phi_3 & \; L_2^2 
       \end{array} 
       \\ 
 \right).
\end{equation} 
We should note that $g_{ab}$ is not exactly identical with the induced metric of the co-ordinate mapping $\Psi$ from ${\it \Delta}$ to ${\bf R}^D\left(D\!=\!2\right)$ but $g_{ab}$ is close to the induced metric of $\Psi$. In fact, $g_{ab}$ is just identical with the induced metric of $\Psi$ if $\Phi\!=\!\Phi_0$. We note also that $g_{ab}$ is independent of $X$, and it depends only on $L$ and $\varphi$; both $L$ and $\varphi$ are considered as functions on ${\it \Delta}$. If $g_{ab}$ is given by the induced metric of $X$ such that  $g_{ab}\!=\!\partial_a X^\mu \partial_b X^\mu$, then ${\it \Delta} \left(\subset M\right)$ can be considered as a subspace of ${\bf R}^d\left(d\!=\!2\right)$, and $L$ and $\Phi$ are identified with $\ell$ and $\phi$ respectively, and the area $A_{\it \Delta}$ is just identical to the area of $X({\it \Delta})$. In this case, the length scale of $M$ is automatically determined by the Euclidean length scale of ${\bf R}^d\left(d\!=\!2\right)$, and hence the length scale of $M$ remains unchanged no matter how $X({\it \Delta})$ and as a consequence $g$ transforms its shape according to the variation of $X$. We note also that $g_{ab}$ in Eq. (\ref{induced_metric}) makes the form $ds^2\!=\!g_{ab}dx_adx_b$ positive definite.

The partial derivatives $\partial_a X^\mu$ are replaced by $X_2\!-\!X_1$ for $a\!=\!1$ and $X_3\!-\!X_1$ for $a\!=\!2$, where $X_1$ and $X_2$ are two terminal vertices of the bond $\ell_1$ in the triangle shown in Fig. \ref{fig-1}(c), and $X_1$ and $X_3$ are those corresponding to the bond $\ell_2$. Since $\int\sqrt{g}d^2x$ corresponds to the area of $M$, $\int\sqrt{g}d^2x$ can be replaced by $\sum_{\it \Delta} A_{\it \Delta} $. Thus, we obtain $S_1\!=\!\sum_{\it \Delta}\left(1/4A_{\it \Delta}\right) \left(L_1^2\ell_2^2\!+\!L_2^2\ell_1^2\!-\!2\ell_1\ell_2\cos\phi_3L_1L_2\cos\Phi_3\right)$. Including the terms, which are the cyclic permutations such that $1\!\to\!2$, $2\!\to\!3$, and $3\!\to\!1$, and using the multiplicative factor $1/3$, we have the expression of $S_1$ in Eqs. (\ref{Disc-Eneg-model-12}) and (\ref{Disc-Eneg-S-1}). This symmetrization  makes $S_1$ reparametrization invariant in the sense that three pairs of edges $(L_1,L_2)$, $(L_2,L_3)$, and $(L_3,L_1)$ can be considered as the coordinate axes.  

If the variable $L$ and $\Phi$ in $S_1$ of Eq. (\ref{Disc-Eneg-model-12}) are replaced by $\ell$ and $\phi$ respectively, then $S_1(\ell,\phi)$ coincides with twice the area of $X(M)$ as a subspace of ${\bf R}^d\left(d\!=\!2\right)$; $S_1(\ell,\phi)\!=\!2\sum_{\it \Delta}a_{\it \Delta}$, where $a_{\it \Delta}$ is the area of $X({\it \Delta})$. If $S_1$ in Eq. (\ref{cont_S1}) is multiplied by the factor $1/2$ such that $(1/2)S_1$, then the factor $1/12$ of $S_1$ in Eq. (\ref{Disc-Eneg-model-12}) is replaced by $1/24$, and as a consequence the corresponding discrete $S_1(\ell,\phi)$ is just identical to the area of $X(M)$. We should note also that the phase structure of the model is independent of the multiplicative factor such as $1/2$ of $S_1$.

The continuous bending energy $S_2$ is given by
\begin{equation} 
\label{cont_S2}
S_2=\frac{1}{2}\int \sqrt{g}d^2x  g^{ab} \partial_a n^\mu \partial_b n^\mu, 
\end{equation} 
where $n^\mu$ is a unit normal vector of the continuous surface in ${\bf R}^d\left(d\!=\!2\right)$ and has values in ${\bf Z}_2\!=\!\{1,-1\}$; the symbol $\mu$ of $n^\mu$ can be dropped. We should note that the expression of $S_2$ of Eq. (\ref{cont_S2}) coincides with that of Polyakov's extrinsic curvature term  $(1/2)\int \sqrt{g}d^2x  K^b_a K^a_b$ in \cite{POLYAKOV-NPB1986} if $d\!=\!3$ and $g_{ab}$ is assumed to be the induced metric such that $g_{ab}\!=\! \partial_a X^\mu \partial_bX^\mu$, where $K_{ab}$ is the second fundamental form defined by $K_{ab}\!=\!-\partial_a X^\mu \partial_b n^\mu$. In fact, it is straightforward to see this by using the equality $\partial_a n^\mu=-K_a^b\partial_b X^\mu$.   We should note that $S_2$ of Eq. (\ref{cont_S2}) is not always identical to the Polyakov's extrinsic curvature term, because $g_{ab}$ in Eq.(\ref{induced_metric}) is different from the induced metric of $X$.

The normal vector $n^\mu$ is defined on the triangles, and therefore we replace $\partial_1 n^\mu$ and $\partial_2 n^\mu$ such that $\partial_1 n^\mu\!\to\!{\bf n}_0\!-\!{\bf n}_2$ and $\partial_2 n^\mu\!\to\!{\bf n}_0\!-\!{\bf n}_1$, where ${\bf n}_i$ are shown in Fig. \ref{fig-2}(b). The discrete version of $S_2$ is then given by
 $ \sum_{\it \Delta} (1/4A_{\it \Delta}) [L_1^2(1\!-\!{\bf n}_0\!\cdot\!{\bf n}_1)\!+\!L_2^2(1\!-\!{\bf n}_0\!\cdot\!{\bf n}_2)
\!-\!L_1L_2\cos\Phi_3({\bf n}_1\!-\!{\bf n}_0)\!\cdot\!({\bf n}_2\!-\!{\bf n}_0)  ]$. Symmetrizing this term by the cyclic permutations just like in the case of $S_1$,  we have $S_2=\!(1/12)\sum_{\it \Delta} S_2({\it \Delta})/A_{\it \Delta}$, where $S_2({\it \Delta})$ is given by Eq. (\ref{Disc-Eneg-S-2}). 

We should comment on why $S_2$ in Eq. (\ref{cont_S2}) is multiplied by the factor $1/2$, which makes the factor of the discrete $S_2$ in Eq. (\ref{Disc-Eneg-S-2}) as $1/12$. This is because $S_2$ in Eq. (\ref{Disc-Eneg-S-2}) is convenient to compare the results with those obtained from the conventional model, whose definition will be mentioned below in Section \ref{Results}. Because of the factor $1/12$, the value of $S_2/N_B$ is comparable to the one of the conventional model at the same $b$.  

The continuous partition function is expressed by 
\begin{equation} 
\label{cont_part_funct}
Z(b)=\int { D}g \int { D}X \exp\left[-S(X,g)\right],
\end{equation} 
where $\int { D}X$ denotes the sum over mappings from $M$ to ${\bf R}^d$, and  $\int { D}g$ the sum over metrics on $M$. The integrations $\int { D}X$ and $\int { D}g$ are considered to represent the sum over surfaces in the string model context. However, we are not going into details of the measures \cite{GinspargMoore-TASI1992}. In this paper, $\int { D}X$ and  $\int { D}g$ are simply replaced by the three-dimensional multiple integrations $\int^\prime \prod _{i=1}^{N} d X_i$ in Eq. (\ref{Part-Func}) and $\int [d L]\int [d \varphi]$ in Eqs. (\ref{measure_1}) and (\ref{measure_2}), respectively. 

\section{Monte Carlo technique}\label{MC_technique}
The vertex position $X\left(\in {\bf R}^d\left(d\!=\!2\right)\right)$ is moved to a new position $X^\prime\!=\!X\!+\!\delta X$, where $\delta X$ is chosen randomly in a small sphere. The radius of the small sphere is fixed as an input parameter of the simulations. The edge length $L$ and the deficit angle $\varphi$ are also changed to $L^\prime\!=\!L\!+\!\delta L$ with $L^\prime>0$ and $\varphi^\prime\!=\!\varphi\!+\!\delta \varphi$, where $\delta L$ and $\delta \varphi$ are chosen randomly in small one-dimensional ranges $\left[-L_0,L_0 \right]$ and  $\left[-\varphi_0,\varphi_0 \right]$. The constants $L_0$ and $\varphi_0$ are fixed to $L_0\!=\!0.5$ and  $\varphi_0\!=\!0.25$. Triangle equalities of Eq. (\ref{triangle-rel}) are also imposed on the variation of $L$.

Both of the energies $S_1$ and $S_2$ vary not only with $\delta X$ but also with $\delta L$ and  $\delta \varphi$, because the area $A_{\it \Delta}$ of the triangle ${\it \Delta}$ depends on edge length $L$ and the internal angle $\Phi$ at least. The internal angle $\Phi$ of ${\it \Delta}$ is also dependent on the deficit angle $\varphi$. Three variables $\varphi$, $L$, and $X$ are assumed to be independently varied; a variation of one variable does not change the remaining variables. Moreover, the intrinsic variables $\varphi$ and $L$ of one triangle are assumed to be independent of those of the other triangles in $M$. Only constraint on the triangles is the fact that $L$ is a common variable shared by two neighboring triangles. These assumptions are understood as reasonable because the variables can be considered as functions on ${\it \Delta}\subset M$. We should note that the assumptions are illegal if $g$ is given by the induced metric $g_{ab}\!=\!\partial_aX^\mu\partial_bX^\mu$, where $M$ is considered to be included in ${\bf R}^d\left(d\!=\!2\right)$, and in this case the bond length of linear triangles varies when the vertex position varies, and as a consequence the length scale of $M$ is fixed to the Euclidean length scale of the bulk space ${\bf R}^d\left(d\!=\!2\right)$ as mentioned in the previous section. 

The triangulated surfaces are obtained from the surfaces constructed in ${\bf R}^d\left(d\!=\!3\right)$ by assuming the third-component of the vertex position as zero. The surfaces in ${\bf R}^d\left(d\!=\!3\right)$ are obtained from the icosahedron by splitting the edges into small pieces and are identical to those used in \cite{KOIB-PRE-2005}.

 One Monte Carlo sweep (MCS) consists of $N$ updates of $X$, $N_B$ updates of $L$, and $N_T$ updates of $\varphi$. Not only the vertices but also the bonds and the triangles are labeled by sequential numbers, and therefore the Metropolis updates are done by using these sequential numbers. The new values of the variables are accepted with the probability ${\rm Min}\left[1,\exp(-\delta S^\prime)\right]$, where $S^\prime$ is the effective Hamiltonian including the terms from the integration measures in Eqs. (\ref{measure_1}) and (\ref{measure_2}) such that $S^\prime\!=\!S_1\!+\!bS_2\!+\!\sum_{i=1}^{N_B} L_i^2\!+\!\sum_{i=1}^{N_T} |\varphi_i|$. 
We have about $50\%$ acceptance rate of $X$, $78\%$ acceptance rate of $L$, and  $93\%$ acceptance rate of $\varphi$. The acceptance rates of $L$ and $\varphi$ are almost independent of variations of $L_0$ and $\varphi_0$.

The bond flip technique is assumed to define the sum over triangulations $\sum_{\rm T}$ in the case of fluid surfaces. Flip of bonds is performed both on $M$ and $X(M)$ in ${\bf R}^d\left(d\!=\!2\right)$. The coordination number $q$ is bounded such that $3\leq q \leq 30$. The phase structure seems not to be so strongly influenced by the assumed upper bound $q_{\rm max}\!=\!30$, because almost all $q$ are smaller than $q_{\rm max}\!=\!30$ in the configurations at relatively small $b$. The length of flipped bond on the surface $X(M)$ is automatically determined by using the canonical coordinate of ${\bf R}^d\left(d\!=\!2\right)$, while the length $L^\prime $ of the flipped bond in $M$ is randomly chosen such that  $L^\prime\!=\!L \!+\!\delta L$, where $L$ is the length of the bond to be flipped and $\delta L(\in \left[-0.5,0.5 \right])$ is a random number. The reason why $L^\prime$ is given by such definition is because we have no information on the length of flipped bond in $M$. Information on $M$ is obtained only locally and limited to a triangle ${\it \Delta}$ and its tangential triangle; the bond length $L$ shared by two triangles only connects one triangle to the other on $M$. This definition of the flipped bond length seems to be a reason why the bond flip strongly influences the equilibrium configurations as mentioned in Section \ref{model}. Thus, we perform the bond flip $N/n_F(n_F\!=\!100)$ times a MCS by choosing bonds randomly. If the flips are performed more frequently; for example $N/n_F(n_F\!=\!2)$ times a MCS, the expected relation $S_1/N\!=\!1$ is considerably violated. This relation is expected due to the scale invariance of the partition function, and no violation is seen in the model defined without the variables $L$ and $\varphi$ even when the flip of bonds is performed more frequently. The rate of acceptance of the bond flip is about $40\%\sim50\%$, which is almost independent of the magnitude of $\delta L$, and the rates of acceptance of the remaining variables are almost the same as in the case of the fixed-connectivity model. 

In order to see the dependence of the results on $N/n_F$, we perform the simulations on the conventional fluid model under $N/n_F\!=\!5$, which is different from the condition $N/n_F(n_F\!=\!100)$. The condition $N/n_F(n_F\!=\!100)$ is given by fixing $n_F$ to be independent of the size $N$, while the condition $N/n_F\!=\!5$ is given by varying $n_F$ to be dependent on $N$ such that $n_F\!=\!N/5$. We find that the critical exponents are almost independent of the conditions. The simulations with $N/n_F\!=\!5$ is relatively time consuming than those with $N/n_F(n_F\!=\!100)$, because the convergence speed is relatively low, or in other words a large number of MCS is necessary to obtain high statistics data, in the simulations with $N/n_F\!=\!5$. This is because the total number of bond flips ($=\!5$) in one MCS is very small in this case. Thus, the condition $N/n_F(n_F\!=\!100)$ is assumed also in model 2. 

The total number of MCS for the production runs after sufficiently large number of MCS for the thermalization is $6\times 10^8\sim 9\times 10^8$ at the transition region of the $N\!=\!2562$, $N\!=\!3612$ surfaces for model 1, and relatively small number of MCS is assumed at the non-transition region and on the smaller surfaces. Almost the same total number of MCS is assumed in the conventional fixed-connectivity model. For model 2 and the conventional fluid model, $6\times 10^8\sim 9\times 10^8$ MCS are assumed at the transition region of the $N\!=\!2562$, $N\!=\!3612$ surfaces, and relatively small number of MCS is assumed at the non-transition region and on the smaller surfaces.

\section{Numerical results}\label{Results}
\subsection{Fixed-connectivity surfaces}
In this subsection, we show results of the fixed-connectivity model, which is model 1, and those of the conventional model defined only by the variable $X\left(\in {\bf R}^d\left(d\!=\!2\right)\right)$. The partition function of the conventional model is given by
\begin{eqnarray}
\label{conventional_fixed_model}
 Z_{\rm c.fix}(b) =  \int^\prime \prod _{i=1}^{N} d X_i \exp\left[-S(X)\right], \quad S=S_1+bS_2,\\
S_1=\sum_{ij}\left(X_i-X_j\right)^2,\quad S_2=\sum_{ij}\left(1-{\bf n}_i\cdot{\bf n}_j\right),
\quad {\bf n}_i\in {\bf Z}_2\!=\!\{1,-1\}, \nonumber
\end{eqnarray}
where $S_1$ and $S_2$ are the Gaussian bond potential and the bending energy, which correspond to $S_1$ and $S_2$ in Eq. (\ref{Disc-Eneg-model-12}). The symbol c.fix in $Z_{\rm c.fix}(b)$ denotes the conventional fixed-connectivity model.

\begin{figure*}[htb]
\centering
\includegraphics[width=14cm]{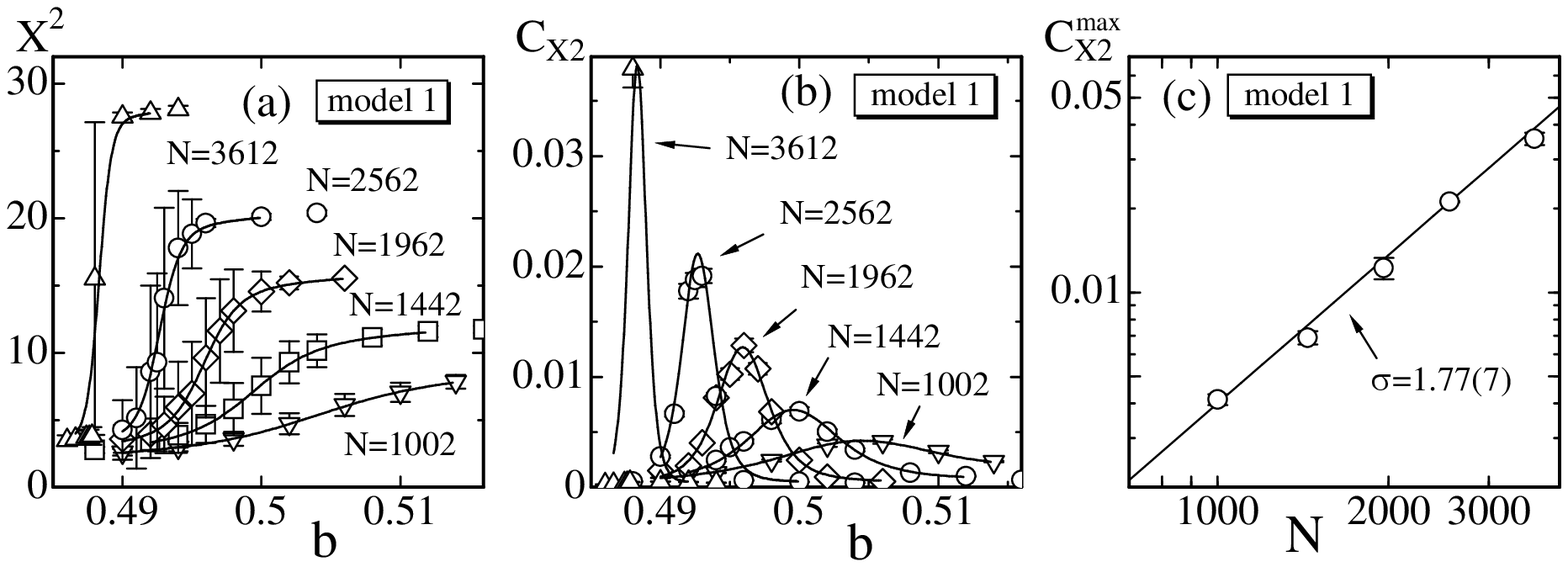}
\caption{ (a) The mean square size $X^2$ vs. $b$ of model 1, (b) the variance $C_{X^2}$ vs. $b$, and (c) the peak values $C_{X^2}^{\rm max}$ vs. $N$ in a log-log scale. The solid curves in (a) and (b) are drawn by the multi-histogram reweighting technique \cite{Janke-histogram-2002}.} 
\label{fig-3}
\end{figure*}
Figure \ref{fig-3}(a) shows the mean square size $X^2$ vs. $b$, where $X^2$ is defined by
\begin{equation}
\label{X2}
X^2={1\over N} \sum_{i=1}^N \left(X_i-\bar X\right)^2, \quad \bar X={1\over N} \sum_{i=1}^N X_i,
\end{equation}
where ${\bar X}$ is the center of mass of the surface.  The solid curves in Figs. \ref{fig-3}(a) and \ref{fig-3}(b) are drawn by the multi-histogram reweighting technique \cite{Janke-histogram-2002}. We see that $X^2$ smoothly varies against $b$, and that $X^2$ rapidly increases with increasing $N$. On the surfaces of $N\!\leq\!2562$, $X^2$ changes up and down many times during the simulations at the transition point. To the contrary, on the largest surface of $N\!=\!3612$, the surface configuration seems to be trapped in one of the two potential minima at the transition point. Thus, the transition is not always correctly reflected on the $N\!=\!3612$ surface.  

The variance $C_{X^2}$ of $X^2$ defined by 
\begin{equation}
\label{CX2}
C_{X^2}={1\over N} \langle \left(X^2-\langle X^2\rangle \right)^2 \rangle
\end{equation}
is plotted in Fig. \ref{fig-3}(b). We find that $C_{X^2}$ has a peak, and the peak value $C_{X^2}^{\rm max}$ increases with increasing $N$. This indicates that the shape transformation is reflected in the fluctuation of $X^2$. 

The peak values $C_{X^2}^{\rm max}$ are plotted in Fig. \ref{fig-3}(c) against $N$ in a log-log scale. The straight line is drawn by fitting the data to $C_{X^2}^{\rm max}\sim N^\sigma$, where $C_{X^2}^{\rm max}$ of the $N\!=\!2562$ surface is the value obtained by the multi-histogram reweighting. Thus, we have 
\begin{equation}
\label{sigma_fixed}
\sigma_{\rm 1} = 1.77\pm 0.07, \quad \left({\rm model\;1}\right). 
\end{equation}
We examined the exponential fitting such that $C_{X^2}^{\rm max}\sim \exp(\sigma)$, however, the power law fitting $C_{X^2}^{\rm max}\sim N^\sigma$ is better than the exponential fitting. This observation can also be seen in all other physical quantities in all of the models, which will be studied in this paper. 

The result $\sigma_{\rm 1} \!=\! 1.77(7)$ in Eq. (\ref{sigma_fixed}) clearly indicates that the transition is of first order.

\begin{figure}[htb]
\centering
\includegraphics[width=14cm]{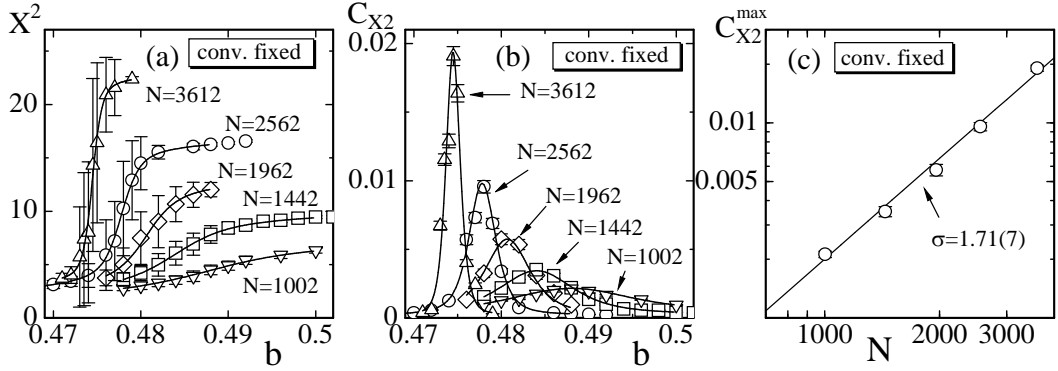}
\caption{ (a) The mean square size $X^2$ vs. $b$, (b) the variance $C_{X^2}$ vs. $b$, and (c) the peak values $C_{X^2}^{\rm max}$ vs. $N$ in a log-log scale. The solid curves in (a) and (b) are drawn by the multi-histogram reweighting technique. The data are obtained by the conventional fixed-connectivity model of Eq. (\ref{conventional_fixed_model}).  } 
\label{fig-4}
\end{figure}
Figures \ref{fig-4}(a)--\ref{fig-4}(c) show the results of the conventional model of fixed connectivity surfaces defined by Eq. (\ref{conventional_fixed_model}). The data shown in the figures correspond to those shown in Figs. \ref{fig-3}(a)--\ref{fig-3}(c). The phase transition is seen in the $N\!=\!3612$ surface of the conventional model in contrast to the case of model 1. In fact, the surface configuration seems not to be trapped in one of the potential minimum states at the transition point; this can be seen in the variation of $X$ against $b$, and for this reason the variance $C_{X^2}$ is correctly computed at the transition region even on  the $N\!=\!3612$ surface. To the contrary, as we see in Figs. \ref{fig-3}(a) and \ref{fig-3}(b) the surface configuration seems to be trapped in the potential minimum states in model 1 on the $N\!=\!3612$ surface. 

\begin{figure}[htb]
\centering
\includegraphics[width=14cm]{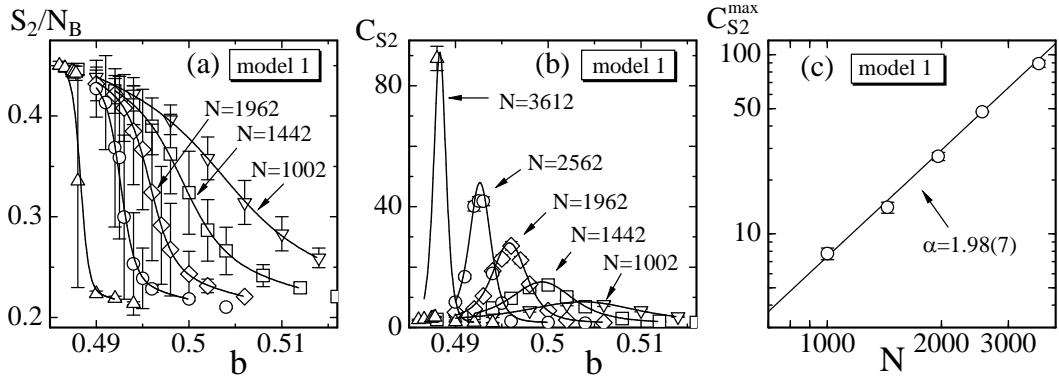}
\caption{ (a) The bending energy $S_2/N_B$ vs. $b$, (b) the specific heat $C_{S_2}$ vs. $b$, and (c) the peak values $C_{S_2}^{\rm max}$ vs. $N$ in a log-log scale.  The data are obtained by model 1 of Eqs. (\ref{Disc-Eneg-model-12})--(\ref{measure_2}). } 
\label{fig-5}
\end{figure}
\begin{figure}[htb]
\centering
\includegraphics[width=14cm]{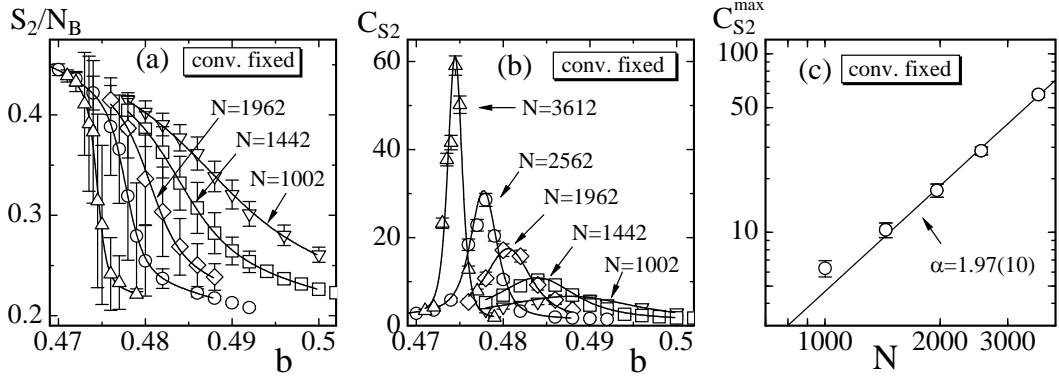}
\caption{ (a) The bending energy $S_2/N_B$ vs. $b$, (b) the specific heat $C_{S_2}$ vs. $b$, and (c) the peak values $C_{S_2}^{\rm max}$ vs. $N$ in a log-log scale. The fitting was done by using the largest four data in (c). The data are obtained by the conventional fixed-connectivity model. } 
\label{fig-6}
\end{figure}
The bending energy $S_2/N_B$ and the specific heat $C_{S_2}$ are plotted in Figs. \ref{fig-5} and \ref{fig-6}, where $C_{S_2}$ is defined by
\begin{equation}
\label{specific_heat}
C_{S_2}=\frac{b^2}{N} \langle \left(S_2-\langle S_2\rangle \right)^2 \rangle.
\end{equation}
The results of model 1 are shown in Fig. \ref{fig-5}, and those of the conventional model are shown in Figs. \ref{fig-6}. The data $C_{S_2}^{\rm max}$ of the $N\!=\!2562$ surface in Fig. \ref{fig-5}(c) is the result of the multi-histogram reweighting, since $C_{S_2}^{\rm max}$ is slightly smaller than the peak of the solid curve as we see in  Fig. \ref{fig-5}(b). The straight line in Fig. \ref{fig-5}(c) is the fitted one of data to $C_{S_2}^{\rm max}\sim N^\alpha$:
\begin{equation}
\label{alpha_fixed}
\alpha_{\rm 1} = 1.98\pm 0.07, \quad \left({\rm model\;1}\right). 
\end{equation}
The data $C_{S_2}^{\rm max}$ and the fitted line of the conventional model are shown in Figs. \ref{fig-6}(c). We see that $S_2/N_B$ of model 1 appears to be trapped in one of the two different values on the $N\!=\!3612$ surface at the transition point in contrast to the conventional model, where $S_2/N_B$ changes up and down many times during the simulations on the $N\!=\!3612$ surface at the transition point. However, the phase structure of model 1 is considered to be identical with that of the conventional model, because the value of $\alpha_1$ of model 1 is almost exactly identical with $\alpha$ of the conventional model.

\begin{figure}[htb]
\centering
\includegraphics[width=13.0cm]{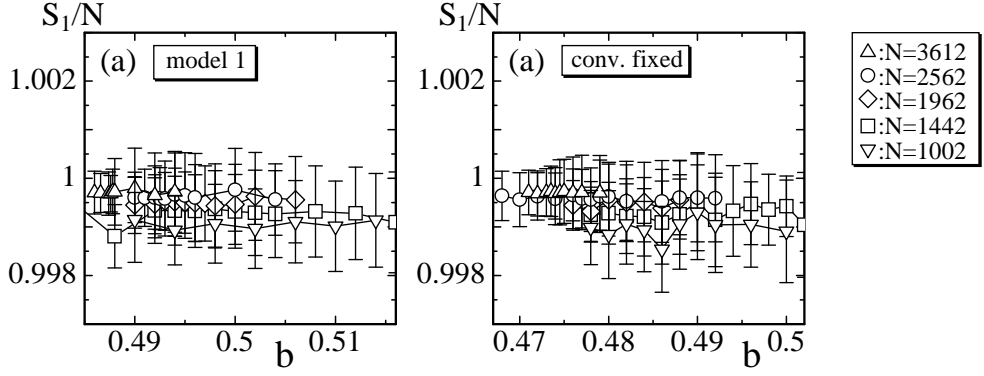}
\caption{The Gaussian bond potential $S_1/N$ vs. $b$ of (a) model 1 of Eqs. (\ref{Disc-Eneg-model-12})--(\ref{measure_2}) and (b) the conventional fixed connectivity model of Eq. (\ref{conventional_fixed_model}).} 
\label{fig-7}
\end{figure}
Finally in this subsection, we plot the Gaussian bond potential $S_1/N$ vs. $b$ in Figs. \ref{fig-7}(a) and \ref{fig-7}(b). From the scale invariance of the partition function, $S_1/N$ is expected to be $S_1/N\!=\!(N-1)/N\!\simeq\!1$. We see from the results in  Figs. \ref{fig-7}(a) and \ref{fig-7}(b) that this expectation is satisfied in both models.   

\subsection{Fluid surfaces}
The fluid surface model denoted by model 2 is defined by the partition function of Eq. (\ref{Part-Func-dyna}), which includes the sum over triangulations. Flips of bond discontinuously change the surface configuration in contrast to the cases of continuous variations of $X$, $L$ and $\varphi$, and hence, the bond flip is performed only $N/n_F(n_F\!=\!100)$ times a MCS as described in Section \ref{MC_technique}. The results are compared with those of the conventional fluid model defined by the partition function 
\begin{equation}
\label{conventional_fluid_model}
 Z_{\rm c.flu}(b) =  \sum_{T} \int^\prime \prod _{i=1}^{N} d X_i \exp\left[-S(X)\right], \quad S=S_1+bS_2,
\end{equation}
where $Z_{\rm c.flu}$ denotes the partition function of the conventional fluid surface model, and $S_1$ and $S_2$ are given by Eq. (\ref{conventional_fixed_model}). The total number of bond flips $N/n_F$ per one MCS is also assumed to be $N/n_F(n_F\!=\!100)$ in the conventional fluid model. The length of flipped bond is automatically obtained in the conventional model, and hence it seems not necessary to reduce $N/n_F$ so small, because the equilibrium surface configuration is not so strongly influenced by the bond flip in contrast to the case of model 2. However, the same $n_F$ is assumed in the conventional model as that of model 2 in order to compare the results under the same condition. 

\begin{figure}[htb]
\centering
\includegraphics[width=14cm]{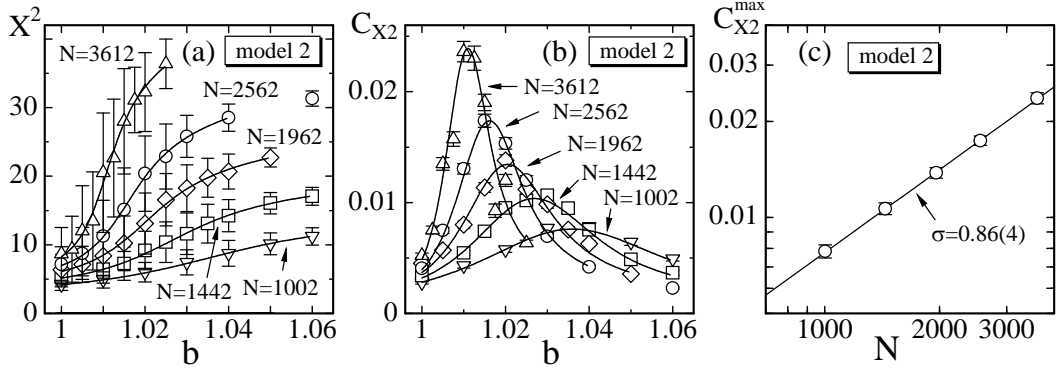}
\caption{(a) The mean square size $X^2$ vs. $b$, (b) the variance $C_{X^2}$ vs. $b$, and (c) the peak values $C_{X^2}^{\rm max}$ vs. $N$ in a log-log scale. The solid curves in (a) and (b) are drawn by the multi-histogram reweighting technique. The data are obtained by model 2.} 
\label{fig-8}
\end{figure}
\begin{figure}[htb]
\centering
\includegraphics[width=14cm]{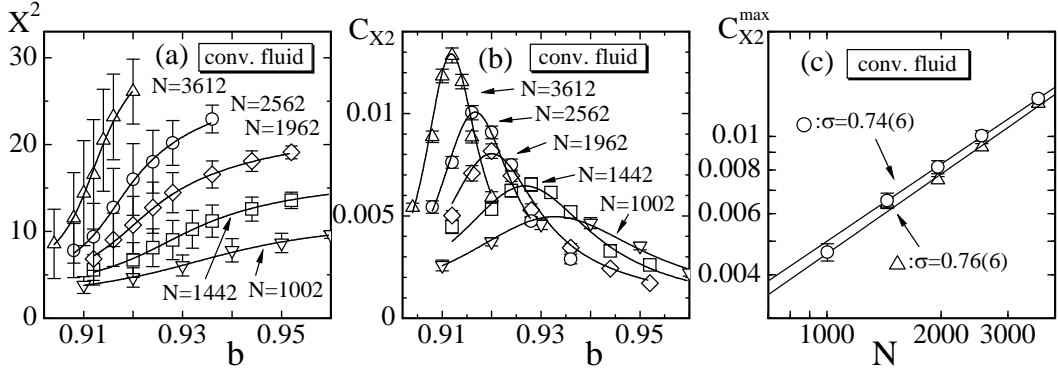}
\caption{(a) The mean square size $X^2$ vs. $b$, (b) the variance $C_{X^2}$ vs. $b$, and (c) the peak values $C_{X^2}^{\rm max}$ vs. $N$ in a log-log scale. The data are obtained by the conventional fluid model of Eq. (\ref{conventional_fluid_model}). The data denoted by symbols $\bigcirc$ and $\triangle$ in (c) correspond to the simulation conditions $N/n_F(n_F\!=\!100)$ and $N/n_F\!=\!5$, respectively. The fitting was done by using the largest four data in (c). } 
\label{fig-9}
\end{figure}
Figures \ref{fig-8} and \ref{fig-9} show the mean square size $X^2$ and the variance $C_{X^2}$ of model 2 and the conventional model. The peak values $C_{X^2}^{\rm max}$ are shown in a log-log scale. We find that the transition is of second-order in both models. The fitted value of the critical exponent, which is defined by $C_{X^2}^{\rm max}\sim N^\sigma$, is
\begin{equation}
\label{sigma_fluid}
\sigma_{\rm 2} = 0.86\pm 0.04, \quad \left({\rm model \; 2}\right).
\end{equation}
The value of $\sigma_{\rm 2}$ is $\sigma_{\rm 2}<1$, and this implies that the transition is of second order, although it is close to first order because $\sigma_{\rm 2}\!\simeq\!1$. Thus, the order of the transition remains unchanged, although the exponent $\sigma_{\rm 2}$ of model 2 is slightly larger than $\sigma_{\rm c.flu.}\!=\!0.74(6)$ of the conventional model. The intrinsic variables $L,\varphi$ slightly strengthen the transition of fluid surface model in contrast to the case of the fixed connectivity model in the previous subsection. In Fig. \ref{fig-9}(c), we show the data denoted by the symbol ($\triangle$), which are obtained with the simulations under the condition $N/n_F\!=\!5$. The exponent $\sigma_{\rm c.flu.}\!=\!0.76(6)$ is almost identical to the $\sigma_{\rm c.flu.}\!=\!0.74(6)$ obtained under the condition $N/n_F(n_F\!=\!100)$, although $C_{X^2}^{\rm max}$ slightly depends on the conditions. This implies that the final results are independent of the simulation condition of $N/n_F$ in the conventional fluid model. Thus, it can also be expected that model 2 is independent of the condition, because $N/n_F(n_F\!=\!100)$ is considered to be sufficiently small.   

\begin{figure}[htb]
\centering
\includegraphics[width=14cm]{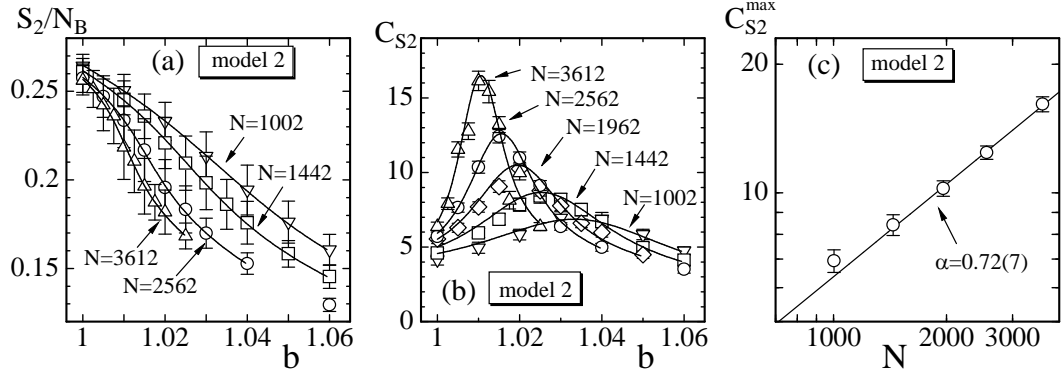}
\caption{(a) The bending energy $S_2/N_B$ vs. $b$, (b) the specific heat $C_{S_2}$ vs. $b$, and (c) the peak values $C_{S_2}^{\rm max}$ vs. $N$ in a log-log scale. The fitting was done by using the largest four data in (c).} 
\label{fig-10}
\end{figure}
\begin{figure}[htb]
\centering
\includegraphics[width=14cm]{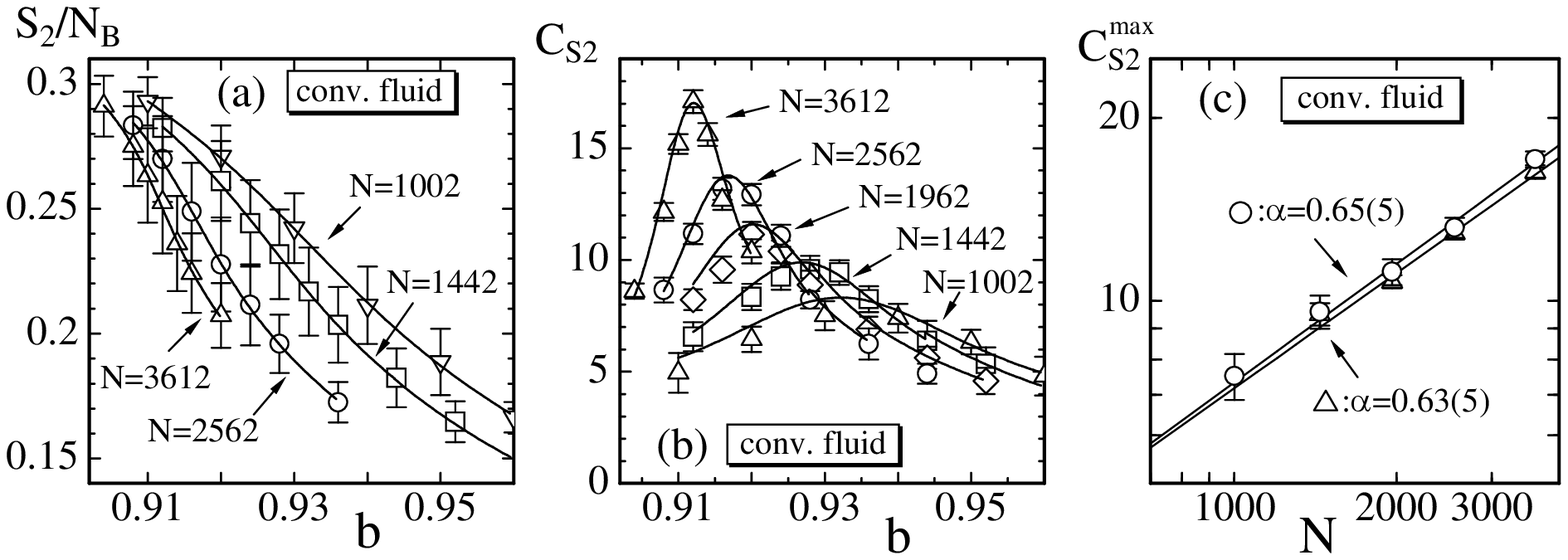}
\caption{(a) The bending energy $S_2/N_B$ vs. $b$, (b) the specific heat $C_{S_2}$ vs. $b$, and (c) the peak values $C_{S_2}^{\rm max}$ vs. $N$ in a log-log scale. The data denoted by symbols $\bigcirc$ and $\triangle$ in (c) correspond to the simulation conditions $N/n_F(n_F\!=\!100)$ and $N/n_F\!=\!5$, respectively.} 
\label{fig-11}
\end{figure}
The bending energy $S_2/N_B$ and the specific heat $C_{S_2}$ of model 2 and the conventional fluid model are shown in Figs. \ref{fig-10} and \ref{fig-11}, where $C_{S_2}$ is defined by Eq. (\ref{specific_heat}). The peak values $C_{S_2}^{\rm max}$ grow larger with increasing $N$, and this behavior of $C_{S_2}^{\rm max}$ of model 2 shown in Fig. \ref{fig-10}(b) is almost identical to that of the conventional model in Fig. \ref{fig-11}(b). $C_{S_2}^{\rm max}$ is plotted in a log-log scale against $N$ in Figs. \ref{fig-10}(c) and \ref{fig-11}(c). The exponent $\alpha$ defined by $C_{S_2}^{\rm max}\sim N^\alpha$, which is the slope of the log-log fit, is given by
\begin{equation}
\label{alpha_fluid}
\alpha_{\rm 2} = 0.72\pm 0.07, \quad \left({\rm model\;2}\right),
\end{equation}
where the fitting was done by using the largest four data. The value of $\alpha_{\rm 2}$ is slightly larger than $\alpha\!=\!0.65(5)$ of the conventional model. This observation is also consistent with the previous ones that the order of the phase transitions of model 1 and model 2 remains unchanged from those of the conventional models. In Fig. \ref{fig-11}(c), we show the data ($\triangle$) obtained with the simulations under the condition $N/n_F\!=\!5$. We find that the exponent $\alpha\!=\!0.63(5)$ is almost identical to $\alpha\!=\!0.65(5)$ obtained under the condition $N/n_F(n_F\!=\!100)$.  

\begin{figure}[htb]
\centering
\includegraphics[width=13cm]{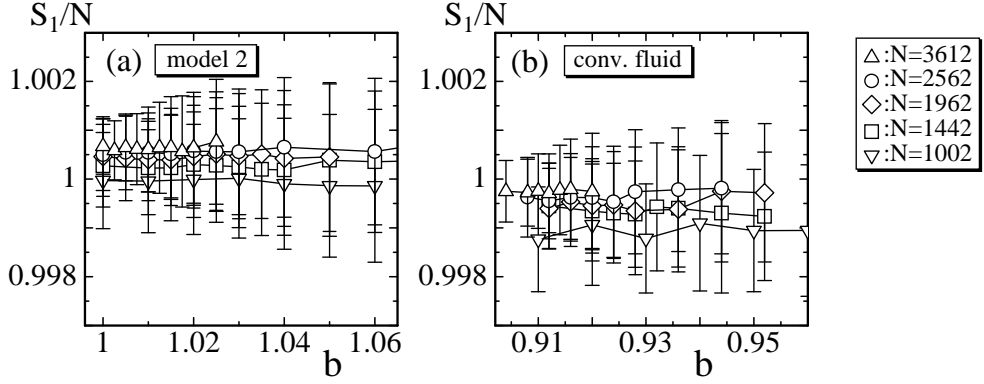}
\caption{The bond potential $S_1/N$ vs. $b$ of (a) model 2 and (b) the conventional fluid model. $S_1/N$ has the expected value $S_1/N\!=\!1$.} 
\label{fig-12}
\end{figure}
Figures \ref{fig-12}(a) and \ref{fig-12}(b) show the bond potential $S_1/N$ vs. $b$. The potential $S_1/N$ in  Fig. \ref{fig-12}(b) is slightly lower than $S_1/N\!=\!1$, this is because $S_1/N\!=\!(N-1)/N$ just as the one in Fig. \ref{fig-7}. In the conventional model, the length of the flipped bond is exactly obtained, because the triangulated surfaces is included in ${\bf R}^2$. Therefore, one can expect that the equilibrium configurations are not so strongly violated by the bond flips. To the contrary, the length of flipped bond in the triangulated surface in $M$ is randomly chosen as described in Section \ref{MC_technique}. Thus, bond flips can influence the equilibrium property of configurations of model 2. This is a reason why $S_1/N$ in  Fig. \ref{fig-12}(a) is slightly larger than $S_1/N\!=\!1$. In fact, the deviation of $S_1/N$ from  $S_1/N\!=\!1$ grows larger when  $N/n_F$ the total number of bond flip per one MCS is assumed to be $N/n_F\!>\!N/100$. If $n_F$ is assumed to be  $n_F\!>\!100$, then we have $S_1/N$ which is more close to $S_1/N\!=\!1$. The deviation of $S_1/N$ seems to grow with increasing $N$ in Fig.\ref{fig-12}(a). This implies that $n_F$ should be increased with increasing $N$.

\begin{figure}[htb]
\centering
\includegraphics[width=10.5cm]{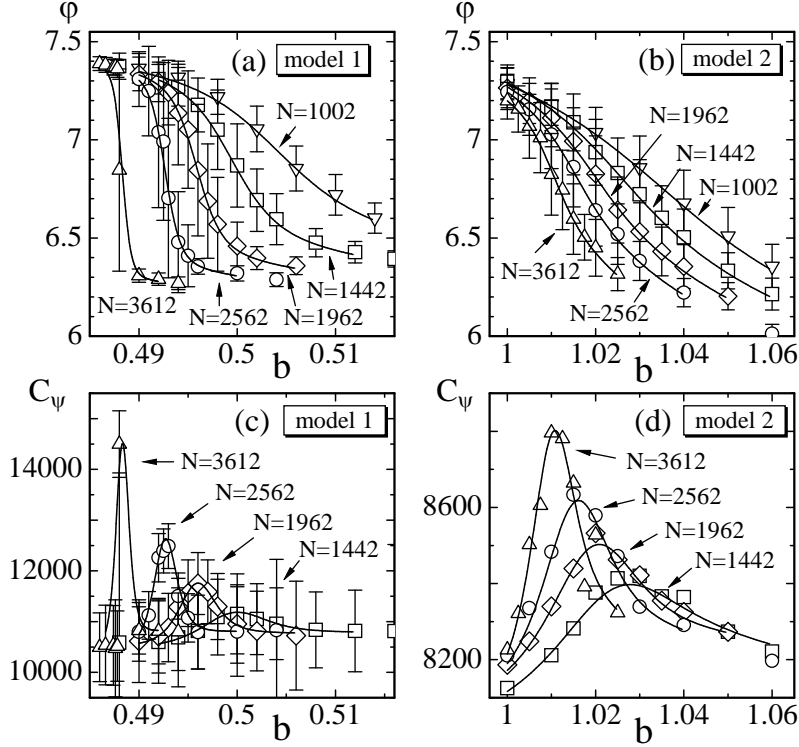}
\caption{The deficit angle $\varphi$ vs. $b$ of (a) model 1 and (b) model 2, where $\varphi$ is defined by $\varphi\!=\!\sum_{i=1}^{N_T}\varphi_i/N_T$, $N_T\!=\!2N\!-\!4$. The unit of $\varphi$ is $[\rm degree]$, which is $[{\rm radian} \times 180/\pi]$. The variance $C_\psi$ is shown in (c) and (d), where no error-bar is shown in (d). The values of $C_\psi(\psi\!=\!N_T\,\varphi)$ in (c) and (d) are reduced by a factor $(\pi/180)^2$ if the unit of $\varphi$ is changed from $[\rm degree]$ to $[{\rm radian]}$. } 
\label{fig-13}
\end{figure}
Finally, we show the deficit angle $\varphi$ vs. $b$ in Figs. \ref{fig-13}(a) and \ref{fig-13}(b). The symbol $\varphi$ denotes $\sum_{i=1}^{N_T}\varphi_i/N_T$, where $\varphi_i$ is defined by Eq. (\ref{deficita}). The variations of $\varphi$ against $b$ are similar to those of $S_2/N_B$ in both model 1 and model 2. The discontinuity seen in $\varphi$ of model 1 is very small compared to the value of $\varphi$ itself, however, we see that the phase transitions are clearly reflected in the internal geometric variables. The variance $C_\psi $ defined by 
\begin{equation}
\label{variance_deficit}
C_\psi ={1\over N} \langle \left(\psi-\langle \psi \rangle \right)^2\rangle, \quad \psi =\sum_{i=1}^{N_T} \varphi_i 
\end{equation}
is plotted in Figs. \ref{fig-13}(c) and \ref{fig-13}(d). The shape of $C_\psi$ is similar to those of $C_{X^2}$ and $C_{S_2}$ in each model, however, the peak value $C_\psi^{\rm max}$ increases only slightly with increasing $N$. For this reason, the scaling of the peak value $C_\psi^{\rm max}$ such as $C_\psi^{\rm max}\sim N^\mu$ is observed neither in model 1 nor in model 2. Although the phase transitions of the models are reflected in $\varphi$, the gap of $\varphi$ is not always considered as a signal of a transition in $M$.

 We should note that $\sum_{i=1}^{N_T}\varphi_i/N_T$ is expected to be zero in the limit of $N\!\to\!0$  on smoothly triangulated surfaces. Thus, non-zero $\varphi$ at $N\!\to\!\infty$ is possible in the model of this paper as mentioned in Section \ref{model}, and in fact it is clear that $\varphi\!>\!0$ in both model 1 and model 2 at the transition points at least. 

We finally comment on the local fluctuation of $\varphi_i$, which is not presented as a figure. In the simulations on both fixed-connectivity and fluid surfaces, the minimum $\varphi_i^{\rm min}$ and the maximum $\varphi_i^{\rm max}$ are respectively comparable to $\pm 2\pi(\sim \pm 3\pi)$, which are considered to be out of the range $-\pi\!<\! \varphi\!<\!2\pi$, which corresponds to $0\!<\! \Phi\!<\!\pi$.  Moreover, the mean value of $|\varphi|$ is about $\pi/3.5$ in both fixed-connectivity and fluid models, and therefore the local fluctuation of $\varphi_i$ is very large compared to the mean value of $\varphi$ estimated from the data in Figs. \ref{fig-13}(a) and \ref{fig-13}(b). The large local fluctuations of $\varphi_i$ seems due to the fact that no interaction of $\varphi_i$ is assumed in the model of this paper. However, we consider that there is no influence of such a relatively large local-fluctuation of $\varphi$ at least on the phase structure as we have confirmed from the presented numerical data.  

\section{Summary and Conclusion}\label{Conclusion}
We have numerically studied a triangulated surface model, which is defined by a mapping $X$ from a two-dimensional spherical surface $M$ to ${\bf R}^d\left(d\!=\!2\right)$. The dynamical variables of the model are the metric $g$ of $M$ and the mapping $X$, which are summed over in the partition function. Hamiltonian $S$ of the model is given by a linear combination of the Polyakov action $S_1$ for strings and the extrinsic curvature $S_2$ such that $S\!=\!S_1\!+\!bS_2$, where $b$ is the bending rigidity. 

By using the Regge calculus technique, the integration over $g$ in the partition function is replaced by the integrations of the edge length $L$ and the deficit angle $\varphi$ of the triangle ${\it \Delta}$ in $M$. The variable $g$ is defined to be a small variation of the induced metric of the coordinate mapping $\Psi$ from ${\it \Delta}$ in $M$ to ${\bf R}^D\left(D\!=\!2\right)$, where the variation is given by the deficit angle $\varphi$. If $\varphi$ is assumed to be  $\varphi\!=\!0$, $g$ is just identical to the induced metric of $\Psi$. In this case, $M$ is still not always completely flat even though ${\it \Delta}$ becomes a linear triangle. Thus, $S$ is defined to be dependent not only on the extrinsic variable $X$ but also on the intrinsic variables $L$ and $\varphi$; $S\!=\!S(X,\{L,\varphi\})$. The integrations of the variable $X$ and the variables $L$ and $\varphi$ are performed in MC simulations by deforming the triangulated surface $X(M)$ in ${\bf R}^d\left(d\!=\!2\right)$ and the triangulated surface $M$, respectively. We should note also that the triangle inequalities are strictly satisfied not only on the triangles in $X(M)$ but also on the ones in $M$ during the MC simulations.

Our attentions are focused on whether the intrinsic variables influence the phase transitions corresponding to the surface fluctuations and the collapse phenomenon. In order to see this influence we study the two variations of the model; the first is the fixed-connectivity model, and the second is the fluid surface model, which is defined on dynamically triangulated lattices. Since the triangle edges on $M$ play a role of local coordinate axes, the dynamical triangulation is considered to make the model reparametrization invariant. The conventional model, which is defined only by using the variable $X$, is also studied in order to compare the results with those of the models in this paper.

Our conclusion is that the internal geometry does not so strongly influence the transition of shape transformation. The order of the transition is of first order in the fixed-connectivity model and of second order in the fluid model. The order of the transition remains unchanged from the corresponding conventional model on both fixed-connectivity and fluid surfaces, although the critical exponents of the transition are slightly different from each other in the case of fluid model. It is also found that the deficit angle $\varphi\left(\!=\!\sum_{i=1}^{N_T}\varphi_i/N_T\right)$ discontinuously changes at the transition at least in the fixed-connectivity model, and hence the transition is reflected in internal geometric variables. 

It is interesting to study the surfaces embedded in ${\bf R}^d\left(d\!=\!3\right)$. It is also interesting to study the case where $g_{ab}$ depends only on the variable $\varphi$, and the case where $g_{ab}$ is not always given by an induced metric such as the one in this paper. These remain to be studied in the future.

\vspace*{3mm}
\noindent
{\bf Acknowledgment}\\

The author H.K. acknowledges an undergraduate student S. Hataoka for his help on computer analyses. The author also acknowledges a referee for helpful comments. 


\end{document}